\documentclass[runningheads]{llncs}
\usepackage{graphicx}
\usepackage{amsmath}
\usepackage{amssymb}
\usepackage{xcolor}
\usepackage{soul}
\usepackage{authblk}
\usepackage{amsmath}
\usepackage[misc]{ifsym}
\usepackage{orcidlink}
\definecolor{lightblue}{RGB}{173,216,230}
\sethlcolor{lightblue}

\newcommand{\mythanks}{These authors contributed equally to this work.}
\begin{document}
\title{StyleFusion TTS: Multimodal Style-control and Enhanced Feature Fusion for Zero-shot Text-to-speech Synthesis}
%
%



\author{Zhiyong Chen\thanks{\mythanks}\orcidlink{0000-0002-9629-6111} \and
Xinnuo Li\textsuperscript{$\star$}\orcidlink{0009-0003-9600-868X} \and
Zhiqi Ai\orcidlink{0009-0005-1034-9972} \and
Shugong Xu\textsuperscript{(\Letter)}\orcidlink{0000-0003-1905-6269}
}

%
%
\institute{School of Communication and Information Engineering, Shanghai University, Shanghai, China\\
\email{\{zhiyongchen,bingpohun,aizhiqi-work,shugong\}@shu.edu.cn}\\ }

\maketitle              
\begin{abstract}
We introduce StyleFusion-TTS, a prompt and/or audio referenced, style- and speaker-controllable, zero-shot text-to-speech (TTS) synthesis system designed to enhance the editability and naturalness of current research literature. We propose a general front-end encoder as a compact and effective module to utilize multimodal inputs—including text prompts, audio references, and speaker timbre references—in a fully zero-shot manner and produce disentangled style and speaker control embeddings. Our novel approach also leverages a hierarchical conformer structure for the fusion of style and speaker control embeddings, aiming to achieve optimal feature fusion within the current advanced TTS architecture. StyleFusion-TTS is evaluated through multiple metrics, both subjectively and objectively. The system shows promising performance across our evaluations, suggesting its potential to contribute to the advancement of the field of zero-shot text-to-speech synthesis. A project website provides detailed information for demonstration and reproduction\footnote{ProjectPage: https://srplplus.github.io/StyleFusionTTS-demo}.

\keywords{Text-to-speech synthesis  \and Voice cloning \and Zero-shot learning \and Multimodal learning.}
\end{abstract}
%
%
%





\section{Introduction}
Text-to-speech (TTS) synthesis has experienced significant advancements in recent years, leading to enhancements in a variety of applications ranging from virtual assistants to accessibility tools. These innovations are exemplified by state-of-the-art expert models \cite{kim2021conditional} and transformer decoder-only models \cite{wang2023neural}. Parallel developments in generative technologies, such as OpenAI's GPT \cite{achiam2023gpt} and AI-generated content (AIGC) using prompt control like StableDiffusion3 \cite{esser2024scaling}, reflect similar progress in adjacent fields.

There is a growing demand for generating audio that can zero-shot mimic the voice timbre of a given reference speaker, while allowing for the customization of content, known as zero-shot TTS (ZS-TTS) or voice cloning \cite{casanova_yourtts_2022}\cite{lee_hierspeech_2023}. This capability significantly enhances system flexibility and scalability.

A major challenge in ZS-TTS systems is the accurate reproduction of the speaker's voice timbre, along with control over speech styles, such as emotion, accent, or characteristics like speed and volume, and maintaining high editability. One method involves using an audio sample for style reference \cite{zhu_multi-speaker_2023}\cite{zhu_metts_2024}, which allows for high customization. However, obtaining such emotional audio can be challenging and may not reliably guide style generation due to the deep entanglement of voice print and stylistic or other acoustic information. Label control is prevalent for style manipulation \cite{qin2023openvoice}, though it often limits variability compared to audio references. Some works utilize prompts to generate speech directly \cite{leng_prompttts_2023}\cite{lyth2024natural}, but this approach compromise the precision of cloning the voice timbre from the speaker's audio.

To address these challenges, we introduce StyleFusion-TTS, an advanced framework designed for zero-shot, style-controlled TTS synthesis. Our methodology combines three input modalities: \textit{text prompts} for natural, interactive dialogue, and/or \textit{style-reference audio} for precise style customization, and \textit{speaker-reference} audio for accurate zero-shot speaker identity cloning. This triple-control-input approach enhances precise control over both the stylistic elements and the distinct voice timbre of the speaker, fully realizing a zero-shot capability in a multi-modal context.

We introduce a compact front-end termed General Style Fusion encoder to encode and disentangle multiple control embeddings for speaker identity and emotions, improving disentanglement for speaker and style modeling. This module facilitates the seamless integration of multi-modal inputs, including text style prompts, audio style references, and speaker voice print or timbre references, all in a fully zero-shot manner. Furthermore, by integrating a novel style control fusion module named HC-TSCM (Hierarchical Conformer Two-Branch Style Control Module) into the state-of-the-art conditional CVAE-based VITS \cite{kong2023vits2} TTS model, ensuring optimal feature fusion and maintain high naturalness in speech synthesis. Our contributions can be summarized as follows:
\begin{itemize}
\item A generalized front-end block capable of representing speaker voice timbre and speech emotional style in a multi-modal and zero-shot manner.
\item An enhanced Hierarchical Conformer Two-Branch Style Control Module (HC-TSCM) that ensures effective feature fusion for zero-shot TTS.
\item The introduction of StyleFusion-TTS, an advancement of existing TTS architecture, designed to produce controllable and natural-sounding speech.
\end{itemize}

\section{Related Work}

The field of style-controllable speech synthesis has seen significant advancements aimed at increasing expressiveness, naturalness, and controllability. Methods such as PromptVC \cite{yao_promptvc_2023} explore voice conversion, while others like Daisy-TTS \cite{chevi_daisy-tts_2024} focus on emotion transfer for single speakers. However, these approaches often fall short in effectively capturing content information or accurately representing speaker identity, primarily concentrating on style conversion and thus restricting their broader applicability.

In multi-speaker text-to-speech (Multi-TTS) synthesis, prompt control has become popular for facilitating natural interaction with human input. Innovations like PromptSpeaker \cite{zhang2023promptspeaker} use prompt information to convey speaker details, while other approaches employ prompts to guide the general style of the speech without disentangling speaker identity and style elements, as seen in Parler-TTS \cite{lyth2024natural}, EmotiVoice \cite{emovoice}, PromptTTS2 \cite{guo2023prompttts}\cite{leng_prompttts_2023}, and PromptStyle \cite{liu2023promptstyle}. MM-TTS systems \cite{guan2024mm} extend this by incorporating visual modalities alongside textual prompts for style guidance. However, the lack of explicit modeling of speaker timbre and the absence of style and speaker disentanglement in these systems restrict their capabilities for precise speaker cloning, limiting their versatility in TTS scenarios.

Zero-shot TTS (ZS-TTS) and voice cloning technologies aim to accurately mimic a speaker's voice print, a critical feature for TTS systems. Notable contributions in this area include VALLE \cite{wang2023neural}, Hierspeech++ \cite{lee_hierspeech_2023}, and StyleTTS2 \cite{li_styletts_nodate}, which focus on precise speaker modeling for effective voice cloning. Additionally, systems like ExpressiveSpeech \cite{zhu_multi-speaker_2023}, Vec-Tok Speech \cite{zhu_vec-tok_2023}, and METTS \cite{zhu_metts_2024} introduce style or emotion control by using reference audio. However, despite their customizability, they often lack the flexibility offered by prompt-based systems. To enhance control flexibility and style robustness, approaches like ZET-Speech \cite{kang_zet-speech_2023} and OpenVoice \cite{qin2023openvoice} employ emotion labels for additional control, though this sometimes restricts user input's ease and expressiveness.

These collective developments underscore a significant evolution towards systems like StyleFusion-TTS, which integrate flexible text prompts and/or audio references for comprehensive style control alongside speaker modeling in zero-shot learning contexts. Leveraging the naturalness of existing ZS-TTS models and incorporating multimodal inputs, StyleFusion-TTS aims to substantially improve speech synthesis customization, enabling more natural and engaging human-computer interactions.

\begin{table}[h]
\caption{Comparison of recent related work on style-controllable ZS-TTS}\label{tab1}
\centering
\resizebox{\linewidth}{!}{
\begin{tabular}{l|c|c|c|c|c}
\hline
\textbf{Methods} & \textbf{ZS Speaker-clone} & \textbf{Prompt Style-control} & \textbf{Audio Style-control} & \textbf{Disentanglement} & \textbf{Ease of Reproduction}\\
\hline
ExpressiveSpeech \cite{zhu_multi-speaker_2023}  & \checkmark &  & \checkmark & & \\
PromptSpeaker \cite{zhang2023promptspeaker}  &  & \checkmark &  &  & \\
PromptStyle \cite{liu_promptstyle_2023} &  & \checkmark &   & \checkmark  & \\
Vec-Tok \cite{zhu_vec-tok_2023} & \checkmark & & \checkmark & & \\
ZET-Speech \cite{kang_zet-speech_2023}  & \checkmark & (Label Only) & & & \\
StyleTTS2 \cite{li_styletts_nodate}  & \checkmark & & \checkmark & & \\
PromptTTS2 \cite{leng_prompttts_2023}  &  & \checkmark & & & \\
METTS \cite{zhu_metts_2024}  & \checkmark & & \checkmark & & \\
ParlerTTS \cite{lyth2024natural}  &  & \checkmark & & & \checkmark \\
\textbf{OpenVoice} \cite{qin2023openvoice}  & \checkmark & (Label Only) & & & \checkmark \\
\textbf{EmotiVoice} \cite{emovoice}  &  & \checkmark &  &  & \checkmark \\
\textbf{MMTTS} \cite{guan2024mm}  &  & \checkmark & \checkmark &  & \checkmark \\
\textbf{StyleFusion-TTS(Ours)}  & \checkmark & \checkmark & \checkmark & \checkmark & \checkmark \\
\hline
\end{tabular}
}
\end{table}

\begin{table}[h]
\caption{Comparison of recent related work on style-controllable ZS-TTS}\label{tab1}
\centering
\resizebox{\linewidth}{!}{
\begin{tabular}{l|c|c|c|c|c}
\hline
\textbf{Methods} & \textbf{ZS Speaker-clone} & \textbf{Prompt Style-control} & \textbf{Audio Style-control} & \textbf{Disentanglement} & \textbf{Ease of Reproduction}\\
\hline
ExpressiveSpeech \cite{zhu_multi-speaker_2023}  & \checkmark &  & \checkmark & & \\
PromptSpeaker \cite{zhang2023promptspeaker}  &  & \checkmark &  &  & \\
PromptStyle \cite{liu_promptstyle_2023} &  & \checkmark &   & \checkmark  & \\
Vec-Tok \cite{zhu_vec-tok_2023} & \checkmark & & \checkmark & & \\
ZET-Speech \cite{kang_zet-speech_2023}  & \checkmark & (Label Only) & & & \\
StyleTTS2 \cite{li_styletts_nodate}  & \checkmark & & \checkmark & & \\
PromptTTS2 \cite{leng_prompttts_2023}  &  & \checkmark & & & \\
METTS \cite{zhu_metts_2024}  & \checkmark & & \checkmark & & \\
ParlerTTS \cite{lyth2024natural}  &  & \checkmark & & & \checkmark \\
OpenVoice \cite{qin2023openvoice}  & \checkmark & (Label Only) & & & \checkmark \\
\textbf{VALL-E} \cite{qin2023openvoice}  & \checkmark & & & & \checkmark \\
\textbf{HierSpeech++} \cite{qin2023openvoice}  & \checkmark &  & & & \checkmark \\
\textbf{NaturalSpeech3} \cite{qin2023openvoice}  & \checkmark &  & & & \checkmark \\
EmotiVoice \cite{emovoice}  &  & \checkmark &  &  & \checkmark \\
MMTTS \cite{guan2024mm}  &  & \checkmark & \checkmark &  & \checkmark \\
\textbf{StyleFusion-TTS(Ours)}  & \checkmark & \checkmark & \checkmark & \checkmark & \checkmark \\
\textbf{TSCM-VITS(Ours)}  & \checkmark &  & & & \checkmark \\
\textbf{StableTTS(Ours)}  & \checkmark &  & & & \checkmark \\
\textbf{StableTTS++(Ours)}  & \checkmark & \checkmark &\checkmark &\checkmark & \checkmark \\
\hline
\end{tabular}
}
\end{table}


\section{StyleFusion-TTS: Multimodal Style and Speaker Control Enhanced TTS}
\begin{figure}
\centering
\includegraphics[width=1.1\linewidth]{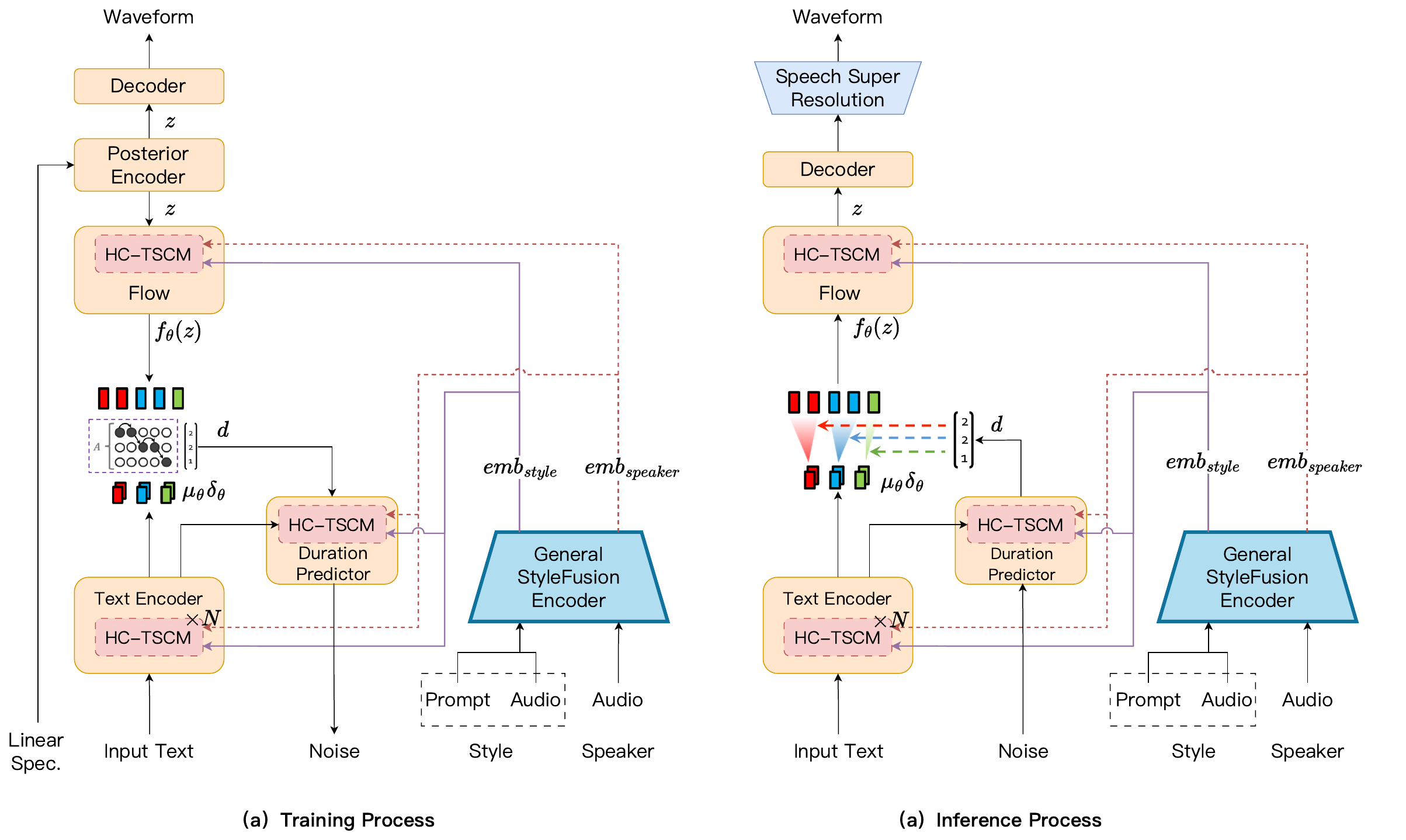}
\caption{Model overview for StyleFusion-TTS} \label{fig_arch}
\end{figure}

Figures~\ref{fig_arch} illustrate the overall architecture of StyleFusion-TTS for training and inference, respectively. The incorporation of the multimodal reference into the VITS architecture \cite{kong2023vits2} aims to enable optimal naturalness and enhanced controllability for speaker and style information. Our model is a flow-conditional-VAE architecture used to synthesize the audio $x$, conditioned on the input text $t$ and control embeddings $c=[emb_{style}, emb_{speaker}]$. The text-to-acoustic distribution is modeled as a normalized flow $f(z)$, which projects acoustic features $z$ to the text features. The optimization of the backbone model involves maximizing the evidence lower bound (ELBO), in alignment with the implementation in original VITS \cite{kim2021conditional}:
\begin{align}
 \log p(x | t,c) \geq \mathbb{E}_{q(f(z)|x,t,c)} & \left[\log p_{\theta}(x|z,t,c)\right] - D_{KL}\left[q(f(z)|x,t,c) \| p(z|t,c)\right] \label{eq:elbo}\\
 L_{syn} &= maximize(ELBO(\cdot)).\label{eq:lsyn}
\end{align}
During training, optimization is conducted with end-to-end training using a HiFi-GAN vocoder and its discriminator. During inference, the output of is upsampled to higher quality using the pretrained SpeechSR module \cite{lee_hierspeech_2023}.

\subsection{General Style Fusion Front-end Encoder (GSF-enc)}
\begin{figure}
\centering
\includegraphics[width=1.05\linewidth]{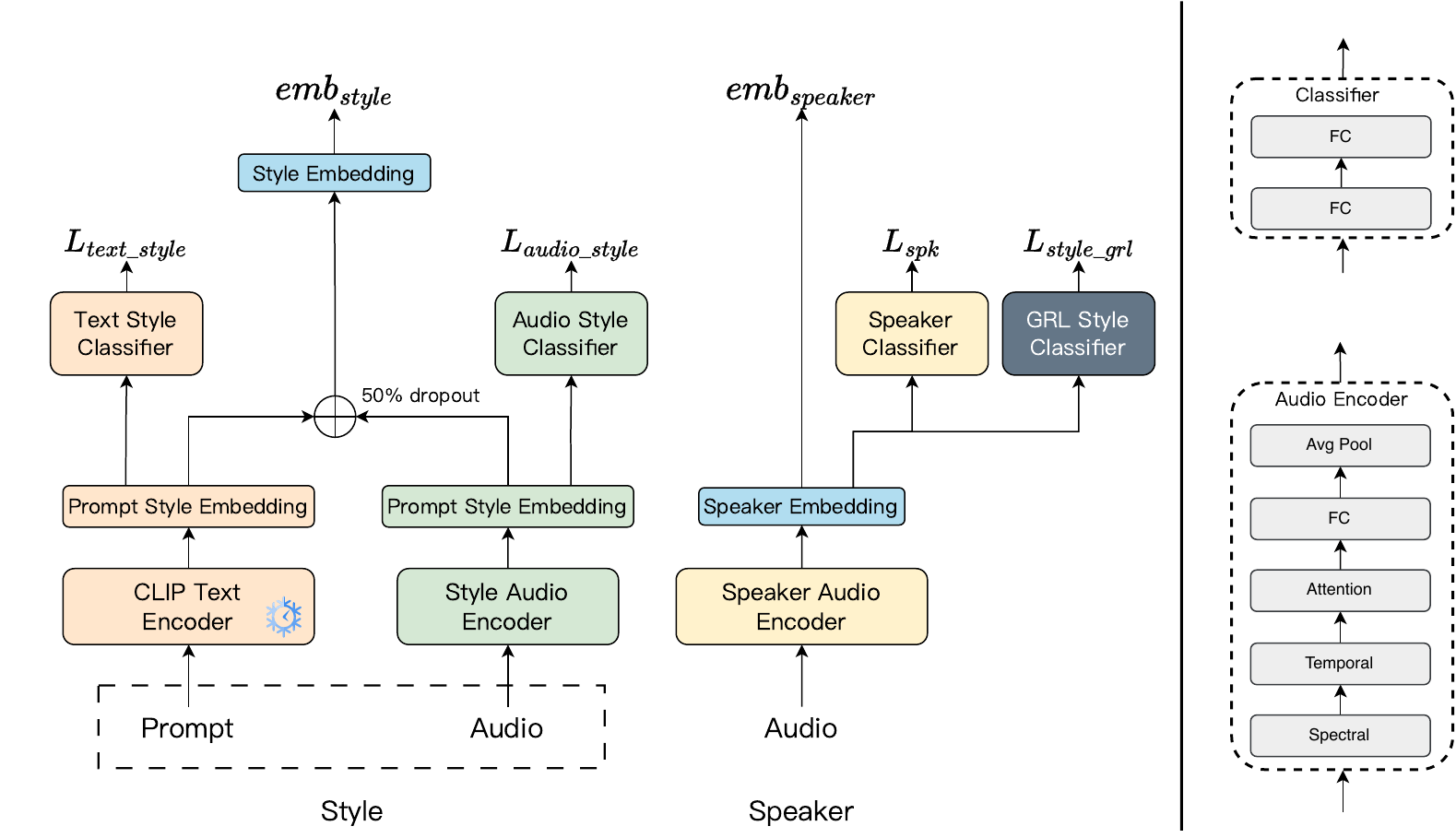}
\caption{Front-end general style fusion encoder (GSF-enc) for speaker and style representation and disentanglement} \label{fig_front}
\end{figure}

\begin{figure}
\centering
\includegraphics[width=\linewidth]{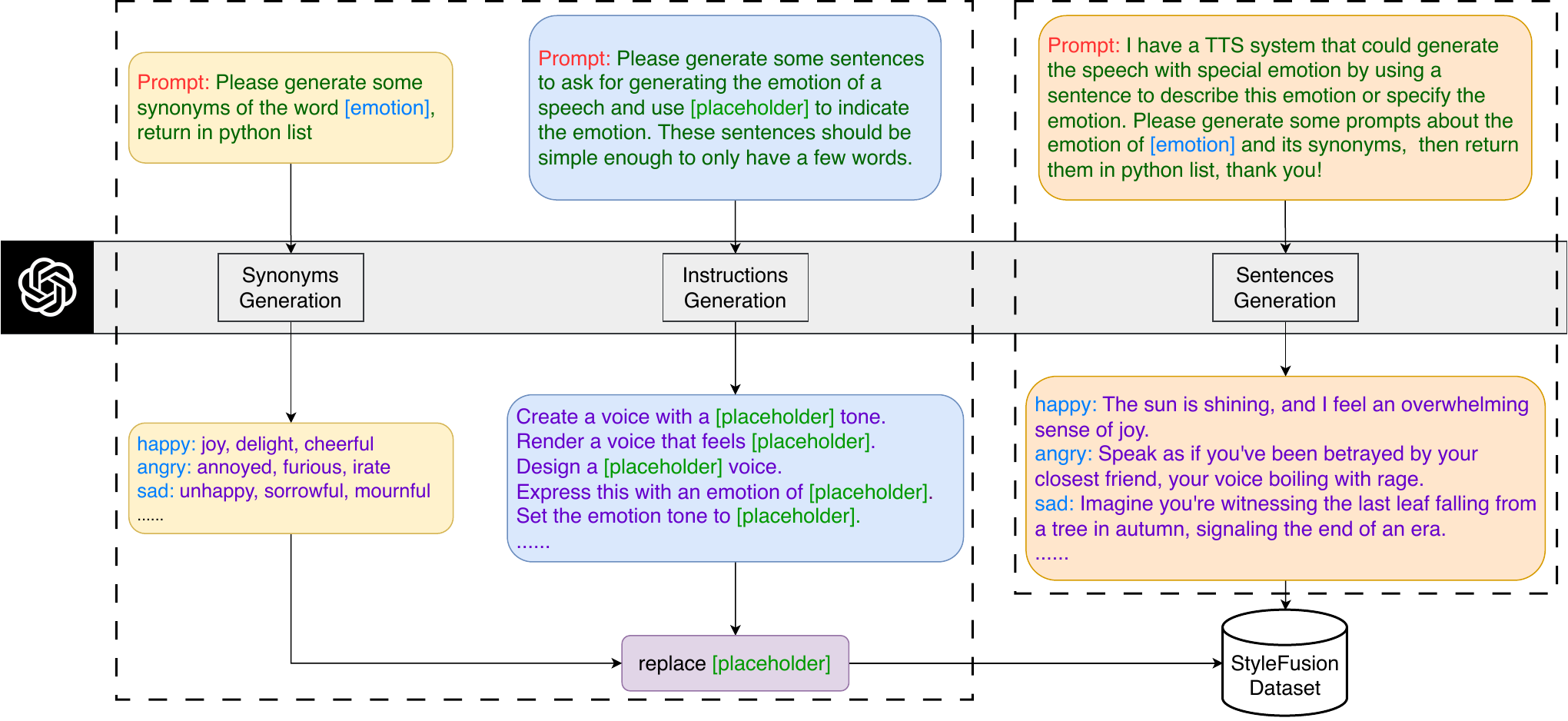}
\caption{Style-control prompt generation pipline with LLM} \label{fig_llm}
\end{figure}

As illustrated in Figure \ref{fig_front}, we propose the General Style Fusion Encoder (GSF-enc) to accurately generate style and speaker embeddings. These embeddings effectively control the style and speaker-related aspects of the backbone models and disentangle these two types of information from multimodal inputs. The text of the input is modeled with the CLIP \cite{radford2021learning} text encoder. The reference text training data is augmented with OpenAI's \cite{achiam2023gpt} LLM to enhance the training text prompts for the style labels in the training datasets. The main procedures are illustrated in Figure \ref{fig_llm}. This process includes generating synonyms for the styles in the dataset to produce multiple keywords (figure left), generating instructions with keywords from the generated keywords (figure middle), and instructing the LLM to directly generate sentences describing each style in the dataset to further augment data variety (figure right). For reference audio, including audio for speaker cloning and style guiding, the audio linear spectrum is extracted as front-end features.

For optimizing the GSF-enc, the $L_{text\_style}$ and $L_{audio\_style}$ jointly supervise the training of the style embedding space output. The $L_{spk}$ supervises the training of the speaker identity, modeling the speaker embedding space and generating the speaker embedding $emb_{speaker}$. While the multimodal inputs are entangled with multiple speech information \cite{ju2024naturalspeech}, a Gradient Reversal Layer (GRL) is employed, supervised with $L_{style\_grl}$, to disentangle the embedding spaces of the speaker and style embeddings. This ensures that the representations of the speaker and style are separated. These components constitute the front-end encoder loss, represented as $L_{GSFenc}$. Following the design pattern noted in related tasks \cite{lee23d_interspeech}, the prompt (text modality) provides a more stable and coarse guidance, while the audio serves as a supplementary reference for better customization of style. To facilitate and/or feature these two style control inputs, a dropout mechanism for generating the final style embedding $emb_{style}$ is used. This mechanism combines the audio emotion embedding $emb_{style\_audio}$ and the prompt emotion embedding $emb_{style\_prompt}$ during training. Therefore:
\begin{align}
&emb_{style} = p_{drop} \cdot emb_{style\_audio}+emb_{style\_prompt}\\
&L_{GSFenc} = L_{text\_style} + L_{audio\_style} + L_{spk} + L_{style\_grl}\\
&L_{total} = L_{GSFenc} + L_{syn}.
\end{align}
The core learning target for StyleFusion-TTS consists of training the $L_{GSFenc}$, and the total loss $L_{total}$ for optimization is the combination of the front-end encoder loss  and the original speech synthesis loss as defined in Equation (\ref{eq:lsyn}).
\subsection{Control-fusion with Hierachical Conformer TSCM (HC-TSCM)}\label{sec:HCTSCM}
\begin{figure}
\centering
\includegraphics[width=\linewidth]{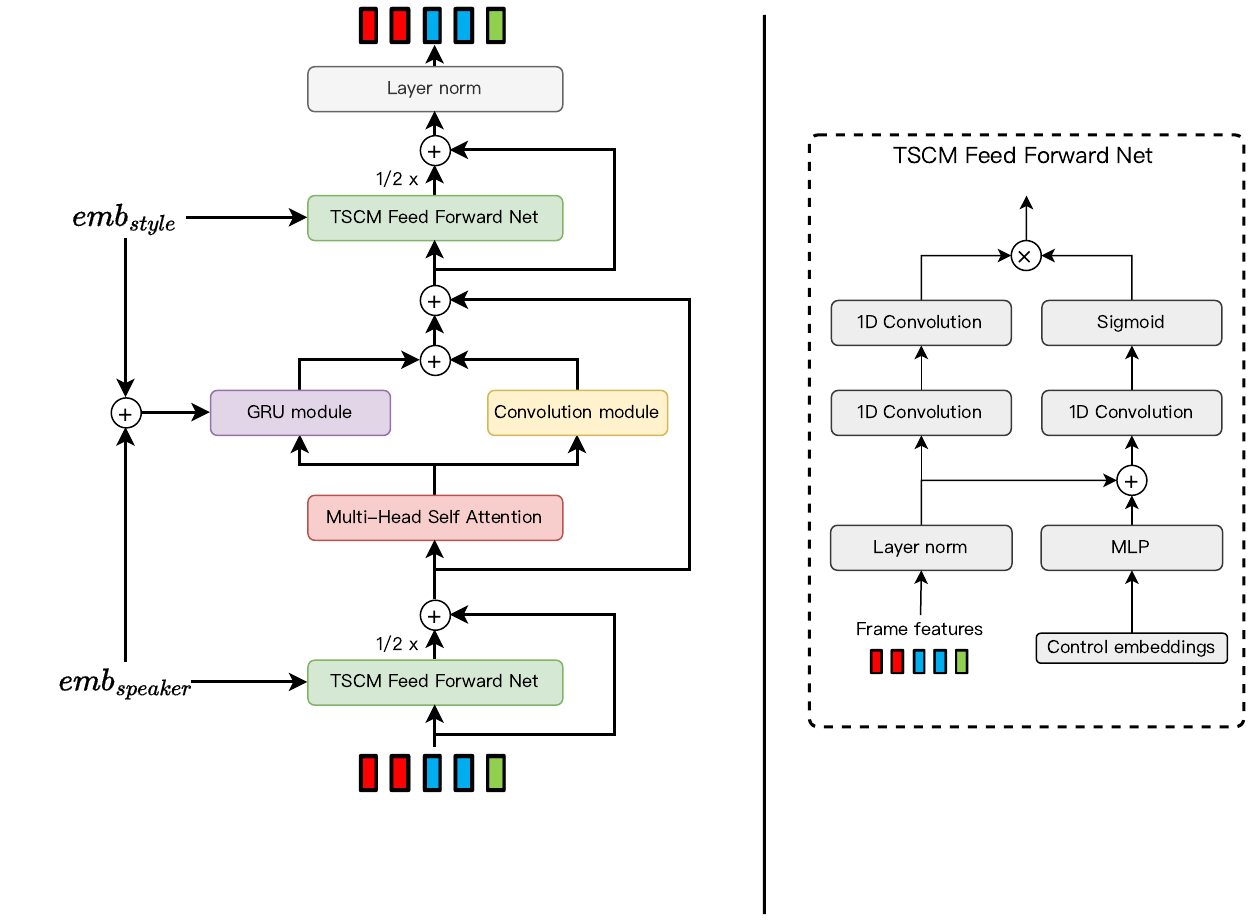}
\caption{Hierachical conformer TSCM (HC-TSCM) for control-fusion} \label{fig_tscm_con}
\end{figure}

The Two-branch Style Control Module (TSCM) \cite{tscm-tts24} is proposed as a fusion method used for advanced style or speaker control, significantly improving speech naturalness for TTS models. We propose using the Hierarchical Conformer TSCM (HC-TSCM), a significant upgrade upon TSCM, to accommodate the case of double control embeddings for disentangled speaker and style embeddings. This method fuses the control embeddings from the GSF-enc into optimal positions within the backbone. As shown in Figure \ref{fig_tscm_con}, the fusion method of HC-TSCM takes the input style vector $emb_{style}$ and $emb_{speaker}$, which are hierarchically fused with $\mathbf{w}_{in}$, the frame-level feature inputs fitting all positions in the backbone. The $\mathbf{w}_{out}$ represents the output of the HC-TSCM module:
\begin{align}
&\mathbf{w} = MSA(FFN_{1}(\mathbf{w}_{in}, emb_{speaker}))\\
&\mathbf{w} = GRU(\mathbf{w}, emb_{speaker}+emb_{style}) + Conv(\mathbf{w})\\
&\mathbf{w}_{out} = LN(FFN_{2}(\mathbf{w}, emb_{style})),
\end{align}
combining multi-head attention (MSA), GRU, ConvNet for local and utterance-wide focusing \cite{tscm-tts24}. The HC-TSCM is adapted as an effective fusion strategy that first renders the speaker information, then the style information, in a hierarchical manner, making the style as variants of the intra-class variation within the modeling of each speaker. StyleFusion-TTS uses HC-TSCM with the goal to precisely control the speaker identity and then make style a flexible variation for each speaker, without losing speaker timbre similarity, thereby achieving superior naturalness (as proven in later experimental studies).

\subsection{Control Implementation with Backbone}
The incorporation of HC-TSCM into the VITS backbone aims to enable optimized style and speaker control. This is implemented within its three core modules: the text content encoder, duration predictor, and text-to-acoustic flow. Figures~\ref{fig_arch} illustrate the overall implementation of our model and the integration of HC-TSCM.
\subsubsection{Text Content Encoder} 
Text content for synthesis is highly relevant to different speakers and emotional styles. Therefore, the transformer block in the original work is smoothly substituted with HC-TSCM. For input $h_{content}$ to each of the text transformer blocks, the result is given by:
\begin{equation}
h_{content} = \text{HC-TSCM}(h_{content}, emb_{style}, emb_{spk})
\end{equation}
\subsubsection{Duration Predictor} 
The duration predictor module is designed to predict the optimal alignments for text-content-frames to acoustic-frames, highly related to the content $h_{content}$, style, and speaker. Therefore, we pre-condition the duration input as the combination of the content, style, and speaker, resulting in the procedure (exemplified by the inference stage):
\begin{align}
& h_{dur\_in} = \text{HC-TSCM}(h_{content}, emb_{style}, emb_{spk}) \\
& d_{predict} = \text{Flow}^{-1}_{dur}(h_{dur\_in}, \epsilon),
\end{align}
where $\epsilon$ is sampled from random noise for prediction variability. The control-rendered $h_{dur_in}$ is then input to the duration flow model $\text{Flow}^{-1}{dur}(\cdot)$ to accurately predict the duration $d{predict}$ for each text-content-frame conditioned on the multiple control embeddings.
\subsubsection{Text-to-acuoustic Flow}
The text-to-acoustic normalizing flow \cite{kong2023vits2} is a module designed to project text to speech at the feature level, highly relevant to speaker and style, to generate optimal prosody and timbre acoustic codes to decode for the final waveform. Modeled as $ f(z) = f_{n} \circ f_{n-1} \circ \cdots \circ f_1(z)$ using the residual coupling layers, where $f(z)$ represents the content frames and $z$ represents the acoustic frames. We define the operation of each flow function $f_{i}(z_{0:C})$ for $C$ channels of input features with the HC-TSCM as:
\begin{align}
& m(z_{0:c}), \sigma(z_{0:c}) = \text{HC-TSCM}(z_{0:c}; {emb}_{style}, {emb}_{spk})\\
& z_{0:c} \leftarrow z_{0:c}\\
& z_{c+1:C} \leftarrow m(z_{0:c}) + \sigma(z_{0:c}) \cdot z_{c+1:C}.
\end{align}
HC-TSCM is implemented for generating the projection elements $m[z_{0:c}]$ and $\sigma[z_{0:c}]$ in text-to-acoustic flow, involving style and speaker control for guiding precise and editable speech synthesis.

\section{Experimental Settings}
Our StyleFusion-TTS system was trained on the \textit{ESD} and \textit{EmoDB} \cite{adigwe2018emotional}\cite{zhou2021seen} multi-speaker TTS corpus. For the purpose of zero-shot speaker cloning testing, we randomly selected 12 speakers, comprising an equal number of males and females. The remaining data was used for the training phase.

In our experiments, all utterances are output at a frequency of 48000 Hz. Our proposed models are trained for 1,000,000 steps. We adhere to the protocols described in the original VITS settings \cite{kim2021conditional} for other training considerations, such as losses and data processing strategies.

The evaluation consists of well-established metrics, including subjective evaluations of MOS, Speaker-MOS for speaker similarity, and Emotion-MOS for style adherence. These are conducted with a series of sample utterances and evaluated by 20 guests who assign scores from 0 to 5 on these three metrics. The subjective testing phoneme content and style prompt align with the online demonstration of \cite{guan2024mm} for fair comparison and to evaluate MOS and Emotion-MOS. Note that \cite{guan2024mm} does not perform speaker cloning. For systems that support speaker cloning, we conduct secondary subjective testing while maintaining these phoneme contents and use the kept-out speakers for enrollment to evaluate their MOS and Speaker-MOS. For systems participating in both tests, their MOS scores are averaged.

Objective evaluations include mean square error in spectrum with ground truth (MCD), word error rate for speech recognition (WER), model-based speaker similarity\footnote{https://github.com/resemble-ai/Resemblyzer} (SECS), and model-based emotional style accuracy\footnote{wav2vec2 emotion recognition: https://huggingface.co/ehcalabres/wav2vec2-lg-xlsr-en-speech-emotion-recognition} (EMO-Acc), aligning with \cite{guan2024mm}\cite{wang2023neural}. For evaluating systems that support style control, the same testing set used in the subjective evaluation is employed. For evaluating systems that support speaker cloning, a testing set of 100 utterances is composed, as ground truth is needed for calculating MCD. More implementation details can be found on our project website$^{1}$.
\section{Results}
\subsubsection{Main results and comparison}
For the comparison of our proposed StyleFusion-TTS models with other state-of-the-art (SOTA) or strong baselines, we demonstrate model comparisons and subjective evaluations, as illustrated in Table \ref{tab:sub}. The comparing models can be split into three categories: SOTA systems focusing on zero-shot speaker cloning (ZS-TTS) which are well-studied, systems optimized for style control that cannot clone speakers, and systems that support both. Our models excel in both areas, and our training data usage is efficient, indicating the superior modeling ability of VITS-based models and great potential for scaling in the future.

Compared with VALLE and HierSpeech++, which are spotlight representatives for transformer-decoder based TTS and CVAE based TTS models respectively, our proposed method outperforms these two SOTA ZS-TTS methods. StyleFusion-TTS still excels and furthermore, they lack the ability for further style control. Systems including MM-TTS, and EmotiVoice-based systems, though capable of controlling the emotional style of speech, are not feasible for performing ZS-TTS tasks and perform suboptimally. Although the OpenVoice system supports both speaker cloning and emotional style control, its performance is not optimal.

Similar results are illustrated for objective evaluation, as shown in Tables \ref{tab:obj_1} and \ref{tab:obj_2}, where we group the comparing systems into those feasible to evaluate speaker identity and style control, and ours participate in both. StyleFusion-TTS demonstrates overall better performance, except in WER compared with the SOTA system HierSpeech++, due to its slow prosody and extremely clear pronunciation of words, trained on 100 times the data compared with ours. Many hard words are modeled, which highlights the necessity for us to scale-up in the next step.


\begin{table}[h]
\centering
\caption{Subjective evaluations and model comparisons for our system and strong baselines, with \textit{N/S} indicating not supported features}\label{tab1}
\resizebox{1.05\linewidth}{!}{
\begin{tabular}{l|c|c|c|c|c}
\hline
\textbf{Methods} & \textbf{Training Data} & \textbf{Feasible Control} & \textbf{MOS} $\uparrow$ & \textbf{Speaker-MOS} $\uparrow$ & \textbf{Emotion-MOS} $\uparrow$ \\
\hline
\hline
VALLE \cite{wang2023neural} & LibriLight (60000hrs) & Speaker & 3.74 & 3.35 & \textit{N/S}\\
HierSpeech++ \cite{zhang2023speechtokenizer} & Multiple (2796hrs) & Speaker & 4.00 & 3.68 & \textit{N/S} \\
\hline
MM-StyleSpeech \cite{guan2024mm} & MEAD (40hrs) \cite{kaisiyuan2020mead} & Style & 3.55 & \textit{N/S} &3.60 \\
MM-TTS \cite{guan2024mm} & MEAD (40hrs) \cite{kaisiyuan2020mead} & Style & 3.56 & \textit{N/S} & 3.60\\
EmotiVoice \cite{emovoice} & LibriTTS/HifiTTS(400hrs) & Style & 4.32 & \textit{N/S} & 3.50\\
\hline
OpenVoice \cite{qin2023openvoice} & LibriTTS(360hrs) & Speaker+Style & 4.17 & 3.58 & 3.81 \\
\hline
\hline
StyleFusion T (Ours) & ESD/EmoDB(30hrs) & Speaker+Style & \textbf{4.34} & \textbf{4.19} & \textbf{4.23}\\
StyleFusion A (Ours) & ESD/EmoDB(30hrs) & Speaker+Style & \textbf{4.25}  & \textbf{4.01} & \textbf{4.16} \\
StyleFusion T+A (Ours) & ESD/EmoDB(30hrs) & Speaker+Style & \textbf{4.29} & \textbf{4.19} & \textbf{4.28} \\
\hline
\end{tabular}
}
\label{tab:sub}
\end{table}


\begin{table}[!ht]
    \centering
    \begin{minipage}{0.48\linewidth}
        \centering
        \caption{Objective evaluations for our models and strong baselines feasible for speaker voice cloning (ZS-TTS)}
        \resizebox{\linewidth}{!}{
        \begin{tabular}{l|l|l|l}
            \hline
            \textbf{Methods} &  \textbf{MCD} $\downarrow$ & \textbf{SECS} $\uparrow$ & \textbf{WER \%} $\downarrow$\\
            \hline
            \hline
            VALLE \cite{wang2023neural} & 11.360& 0.707& 14.165\\
            OpenVoice \cite{qin2023openvoice} &  8.562&  0.686&  12.619\\
            HierSpeech++ \cite{zhang2023speechtokenizer} &  10.726&  0.748&   \textbf{6.229}\\
            \hline
            \hline
            StyleFusion T (Ours)&  \textbf{5.775}&  \textbf{0.807}&   14.644\\
            StyleFusion A (Ours)&  \textbf{5.825}&  \textbf{0.795}&   14.134\\
            StyleFusion T+A (Ours)&  \textbf{5.762}&  \textbf{0.810}&   13.960\\
            \hline
        \end{tabular}
        }
        \label{tab:obj_1}
    \end{minipage}%
    \hfill
    \begin{minipage}{0.48\linewidth}
        \centering
        \caption{Objective evaluations for our model and strong baselines for emotional style control}
        \resizebox{\linewidth}{!}{
        \begin{tabular}{l|l|l}
            \hline
            \textbf{Methods} & \textbf{WER \%} $\downarrow$ & \textbf{EMO-Acc \%} $\uparrow$\\
            \hline
            MM-StyleSpeech \cite{guan2024mm} & 19.17& 13 (-37)\\
            MM-TTS \cite{guan2024mm}&  16.12& 6 (-44)\\
            OpenVoice \cite{qin2023openvoice} & \textbf{13.66}& 25 (-25)\\
            EmotiVoice \cite{emovoice}&  22.53&  12 (-37)\\
            \hline
            \hline
            StyleFusion T+A (Ours) &  \textbf{13.96}&  \textbf{50 (0)} \\
            \hline
        \end{tabular}
        }
        \label{tab:obj_2}
    \end{minipage}
\end{table}



\begin{figure}[!h]
\centering
\includegraphics[width=\textwidth]{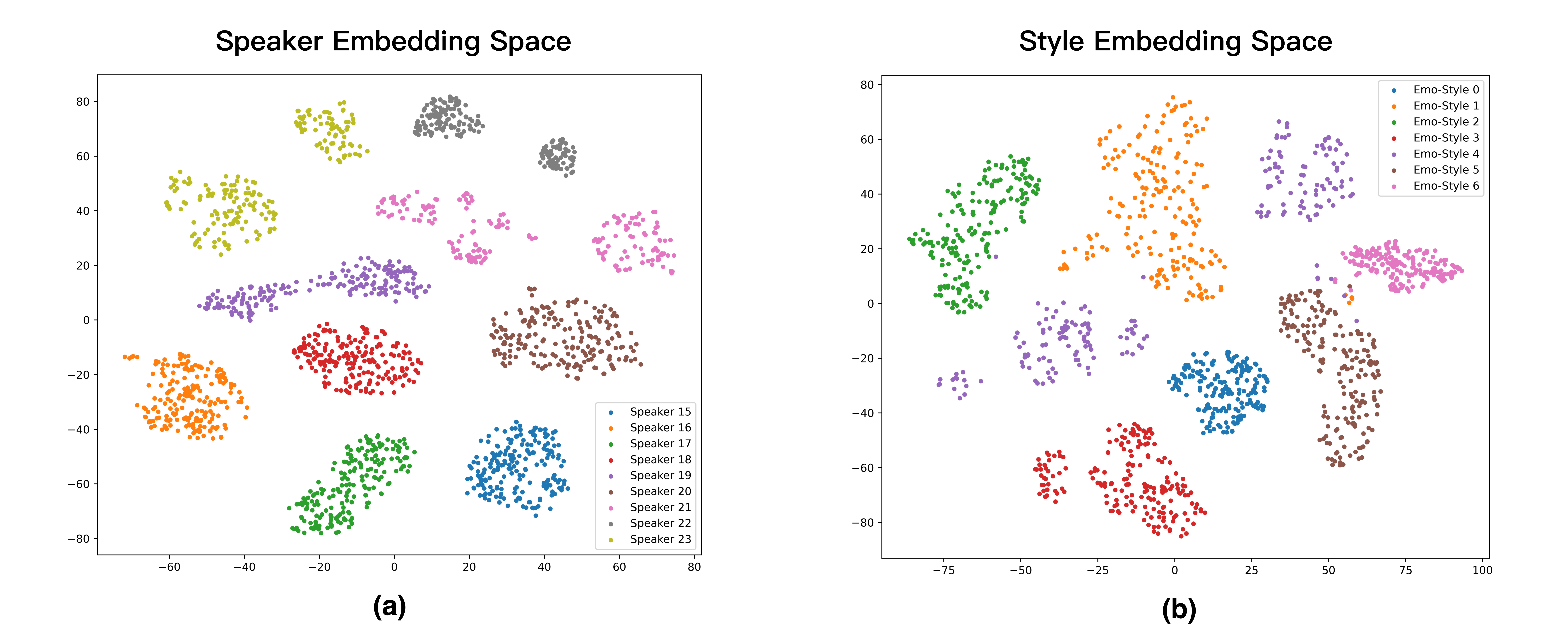}
\caption{(a)-(b) The style and speaker embeddings of GSF-enc} \label{fig:emb_f}
\label{fig:emb}
\end{figure}

\begin{figure}[!h]
\centering
\includegraphics[width=\textwidth]{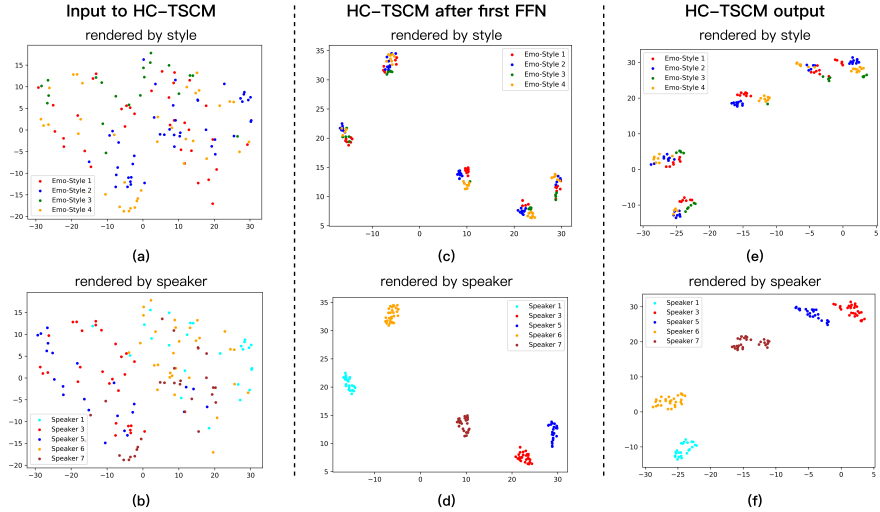}
\caption{(a)-(f) Embedding visualization for HC-TSCM modules at three positions in the \textit{last} acoustic-flow layer, rendered in different colors for various speakers or styles} \label{fig_emb}
\label{fig:emb_hc}
\end{figure}

\begin{table}[!h]
\centering
\caption{Ablation studies comparing models with and without HC-TSCM-based speaker and style-aware fusion}
\resizebox{0.8\linewidth}{!}{
\begin{tabular}{c|c|c|c|c}
\hline
\textbf{Methods} & \textbf{Fusion Modules} & \textbf{MCD} $\downarrow$& \textbf{SECS} $\uparrow$& \textbf{WER \%} $\downarrow$\\
\hline
\hline
\textbf{\textit{StyleFusion}} T+A & HC-TSCM (Ours) & \textbf{5.762} & \textbf{0.810} & \textbf{13.961} \\
\hline
\hline
\textit{w/o HC-TSCM} T & TSCM \cite{tscm-tts24} & 6.007 & 0.737 & 25.950 \\
\textit{w/o HC-TSCM} A & TSCM \cite{tscm-tts24} & 11.069 & 0.550 & 95.963 \\
\textit{w/o HC-TSCM} T+A & TSCM \cite{tscm-tts24} & 6.008 & 0.755 & 19.762 \\
\hline
\textit{w/o TSCM} T  & Naive VITS \cite{wang2023neural} & 6.594& 0.693& 17.857\\
\textit{w/o TSCM} A  & Naive VITS \cite{wang2023neural} & 6.470& 0.714& 17.857\\
\textit{w/o TSCM} T+A  & Naive VITS \cite{wang2023neural} & 6.422& 0.742& 20.358\\
\hline
\end{tabular}
}
\label{tab:abl_hc}
\end{table}


\begin{table}[!h]
\centering
\caption{Study of control effects with contradictory versus consistent text and audio modalities}\label{tab1}
\resizebox{\linewidth}{!}{
\begin{tabular}{l|c}
\hline
\textbf{Style Control} &  \textbf{EMO-Acc \%} $\uparrow$\\
\hline
\hline
Neutral text prompt + Emotional audio prompt  &  63.5\\
Emotional text prompt + Neutral audio prompt  &  77.8\\
Emotional text prompt + Emotional audio prompt  & \underline{83.3}\\
\hline
\hline
Negative emotional text prompt + Positive emotional audio prompt & \underline{36.4}\\
Positive emotional text prompt + Negative emotional audio prompt & \underline{36.0}\\
\hline
\end{tabular}
}
\label{tab:abl_ta}
\end{table}
\subsubsection{Auxiliary studies}
We conducted additional evaluations and analyses to assess the effectiveness of each module we proposed. As illustrated in Table \ref{tab:abl_hc}, using our proposed HC-TSCM for control fusion performs overall better than the baseline TSCM and the naive VITS original control fusion method (i.e., simple concatenation of $emb_{style}$ and $emb_{speaker}$). In Figure \ref{fig:emb_f}, we demonstrated that the embeddings for style and speaker control are effective (a) \& (b) for representing the speakers and styles. For the plots in Figure \ref{fig:emb_hc} (a)-(f), we observe a clearer view of the effectiveness of the HC-TSCM module, using the visualization of the HC-TSCM from the last flow module. Each point represents an utterance-level mean. By hierarchically adding speaker and style information, we observe a transformation from a chaotic distribution at input in (a) \& (b), to a more clustered distribution for speakers in (c) \& (d), and finally clustered features for speakers with intra-speaker distinguishable features for emotional styles in (e) \& (f). These results demonstrate the effective generation of speaker and style control by using the HC-TSCM modul and GSF-enc, highlighting the effectiveness of our design. 

In Table \ref{tab:abl_ta}, we conduct experiments with 100 prompt and audio pairs for each scenario of using prompts and audio, and we evaluate the EMO-Acc for the synthesized speech. This demonstrates that the audio and prompt style control modalities are complementary when both emotions express the same tendency, and if set to be contradictory, the emotional effect will be neutralized, as indicated by the degradation in EMO-Acc metric. This highlights the effective multimodal modeling capability for style control embedding representation of our proposed GSF-enc and its expected traits.


\section{Conclusions}
We proposed StyleFusion-TTS, a prompt and/or audio referenced, style- and speaker-controllable, zero-shot TTS system. By innovatively employing the HC-TSCM fusion module, our system achieves optimal integration of speaker and style control embeddings. The introduction of the general front-end encoder facilitates the effective utilization of multimodal inputs, including both prompt and reference audio, and improves disentanglement. These proposed methods significantly broaden the applicability and flexibility of TTS technologies while maintaining naturalness. Our comprehensive evaluations, with both subjective and objective metrics, confirm the performance of StyleFusion-TTS. Looking ahead, as we continue to refine and expand this framework, StyleFusion-TTS will extend to a multilingual version, enhancing its precision and expressiveness for improved effectiveness.

\section*{Acknowledgements}
This work was supported in part by the National High Quality Program grant TC220H07D, the National Natural Science Foundation of China (NSFC) under Grant 61871262, the National Key R\&D Program of China grants  2022YFB2902000, the Innovation Program of Shanghai Municipal Science and Technology Commission under Grant 20511106603, Foshan Science and Technology Innovation Team Project grant FS0AAKJ919-4402-0060.

%
%

%
%
%

\bibliographystyle{splncs04}
\bibliography{mybibliography}

%





\end{document}